\documentclass[10pt]{report}
\usepackage{array}
\usepackage{graphicx}
\begin{document}
{\bfseries \noindent Vacuum fluctuation force on a rigid
Casimir cavity in a gravitational field}\\
\vspace{0.5cm} $E. \; Calloni^{1,2}$, $L. \; Di \; Fiore^{2}$, $G.
\; Esposito^{2}$,
$L. \; Milano^{1,2}$, $L. \; Rosa^{1,2}$  \\
\vspace{0.5cm}
\noindent 1: Universit\`{a} Federico II di Napoli,
Complesso Universitario di Monte Sant' Angelo, Via Cintia, 80126 Naples \\
2: INFN, Sezione di Napoli, Complesso Universitario di Monte Sant'
Angelo, Via Cintia, 80126 Naples \\
\large \baselineskip=0.8cm
\font\bigbf = cmbx12
\font\bigbigbf = cmbx12 scaled \magstep2
\vspace{1.5cm}
\section*{{\bfseries Abstract}}
\vspace{0.5cm}
We discuss the possibility of verifying the equivalence principle
for the zero-point energy of quantum electrodynamics,
by evaluating the force, produced by vacuum fluctuations, acting on a
rigid Casimir cavity in a weak gravitational field. The resulting
force has opposite direction with respect to the gravitational
acceleration; the order of magnitude for a multi-layer cavity
configuration is derived and experimental feasibility is discussed,
taking into account current technological resources.
\section*{{\bfseries Introduction}}
\vspace{1.5cm}

Although much progress has been made in the evaluation and
experimental verification of effects produced by vacuum energy in
Minkowski space-time [1-6], it remains unclear why the observed
universe exhibits an energy density much smaller than the one
resulting from the application of quantum field theory and the
equivalence principle [7,8]. Various hypotheses have also been put
forward on the interaction of virtual quanta with the gravitational
field. For instance some arguments seem to suggest that virtual photons
do not gravitate [8], while other authors have suggested that
Casimir energy contributes to gravitation [9].
So far it seems fair enough to say that no experimental
verification that vacuum fluctuations can be treated according to
the equivalence principle has been obtained as yet, even though
there are expectations, as we agree, that this should be the case.

Motivated by all these considerations, our paper computes the effect
of a gravitational field on a rigid Casimir cavity, evaluating the
net force acting on it. The order of magnitude
of the resulting force, although not allowing an immediate
experimental verification, turns out to be compatible with the current
extremely sensitive force detectors, actually the interferometric
detectors of gravitational waves. The Casimir cavity is rigid in
that its shape and size remain unchanged under certain particular
external conditions such as, for example, the absence of
accelerations or impulses. We evaluate the force acting on this
non-isolated system at rest in the gravitational field of the
earth by studying the regularized energy-momentum
tensor. The associated force
turns out to have opposite direction with respect to
gravitational acceleration. Orders of magnitude are discussed
bearing in mind the current technological resources as well as
experimental problems, and physical relevance of the analysis
is stressed.

\section*{Evaluation of the force}

In order to evaluate the force due to the gravitational field
let us suppose that the cavity has geometrical configuration of two
parallel plates of proper area $A = L^{2}$ separated by the proper
length $a$. The system of plates is taken to be orthogonal to the
direction of gravitational acceleration $\vec g$.

In classical general relativity, the force density can be
evaluated according to [10]
\begin{equation}
f_{\nu}=-{1\over \sqrt{-{\rm det} \; g}}
{\partial \over \partial x^{\mu}}
\Bigr(\sqrt{-{\rm det} \; g} \; T_{\; \nu}^{\mu}\Bigr)
+{1\over 2}{\partial g_{\rho \sigma}\over \partial x^{\nu}}
T^{\rho \sigma},
\end{equation}
where $T^{\mu \nu}$ is the energy-momentum tensor of matter,
representing the energy densities of all non-gravitational fields,
and gravity couples to $T^{\mu \nu}$ via the Einstein equations:
$R_{\mu \nu}-{1\over 2}g_{\mu \nu}R=8\pi G T_{\mu \nu}$.
However, we are eventually interested in quantum field theory in
curved space-time, where gravitation is still treated
classically, while matter fields are quantized. At this stage the
key assumption is that one can take the expectation value of
the quantum energy-momentum tensor in some state and evaluate
the external force as if it were a classical force. This should
therefore represent the expectation value of the ``quantum force''
in the given state. According to this theoretical model, we have
to find the regularized energy-momentum tensor
$\langle T^{\mu \nu} \rangle$ and insert it into the right-hand
side of Eq. (1) to find eventually $\langle f_{\nu} \rangle$.
If we were in Minkowski space-time with metric
$\eta$, we might exploit the result of Ref. [11],
according to which
\begin{equation}
\langle T^{\mu \nu} \rangle ={\pi^{2}{\hbar}c \over 180 a^{4}}
\left({1\over 4}\eta^{\mu \nu}-{\hat z}^{\mu}{\hat z}^{\nu}
\right),
\end{equation}
where ${\hat z}^{\mu}=(0,0,0,1)$ is the unit spacelike
4-vector orthogonal to the plates' surface. If such an analysis
is performed for an accelerated but locally orthonormal
reference frame in curved space-time, corresponding to our
system at rest in the earth's gravitational field, the Minkowski
line element should be replaced by the line element near the
observer's world line. As shown in detail in Ref. [12] this
reads (neglecting possible rotation effects)
\begin{equation}
ds^{2}=-(1+2A_{j}x^{j})(dx^{0})^{2}+\delta_{jk}dx^{j}dx^{k}
+O_{\alpha \beta}(|x^{j}|^{2})dx^{\alpha}dx^{\beta}.
\end{equation}
Here $c^{2}{\vec A}$
(with components $(0,0,|{\vec g}|)$), the observer's acceleration
with respect to the local freely falling frame,
shows up in the correction
term $-2A_{j}x^{j}$ to $g_{00}$, which is proportional to
distance along the acceleration direction. Note that
first-order corrections to the line element are unaffected
by space-time curvature. Only at second order, which is
beyond our aims, does curvature begin to show up. With this
understanding we find, on setting
\begin{equation}
K \equiv {\pi^{2}{\hbar}c \over 180 a^{4}},
\end{equation}
that the non-vanishing components of the regularized
energy-momentum tensor which contribute to $\langle f_{z}
\rangle$ are
\begin{equation}
\langle T_{\; 3}^{3} \rangle = \langle T^{33} \rangle
=-{3\over 4}K,
\end{equation}
\begin{equation}
\langle T^{00} \rangle =-{K\over 4}{1\over 1+2Az}.
\end{equation}
Since (of course $x^{3} \equiv z$ and hence $\langle f_{3} \rangle$
denotes $\langle f_{z} \rangle$)
\begin{equation}
\langle f_{z} \rangle = -{1\over \sqrt{1+2Az}}{\partial \over
\partial x^{\mu}}\Bigr(\sqrt{1+2Az}T_{\; 3}^{\mu}\Bigr)
+{1\over 2}{\partial g_{\eta \rho}\over \partial z}
T^{\eta \rho}
=\left({3\over 4}K+{1\over 4}K \right){A\over 1+2Az},
\end{equation}
we find for the component of the
full force along the $z$-axis the formula
\begin{equation}
\langle F_{z} \rangle \cong aL^{2} \langle f_{z} \rangle
\cong aL^{2}gK,
\end{equation}
where integration over the volume $V=aL^{2}$ has reduced to
a simple multiplication by $aL^{2}$ to our order of
approximation.

Interestingly, the resulting force has
opposite direction with respect to the gravitational
acceleration. Note also that the force density
in Eq. (7), and hence the force itself in Eq. (8),
is the sum of two terms: the latter, proportional
to $g{K\over 4}$, arises from the Casimir energy
encoded into $T^{00}$, and can be interpreted as the
Newtonian repulsive force (i.e. ``push'') on
an object with negative energy.
The former term, proportional to ${3\over 4}gK$, results from
the pressure along the acceleration axis and can be interpreted
as the mass contribution of the spatial part of the
energy-momentum tensor. With this understanding, our result
agrees with the equivalence principle, according to which for
every pointlike event of space-time, there exists a sufficiently
small neighbourhood such that in every local, freely falling
frame in that neighbourhood, all laws of physics obey the laws
of special relativity (as a corollary, the gravitational
binding energy of a body contributes equally to the inertial mass
and to the passive gravitational mass).
In particular, we agree with the
statement that a system with a given rest energy momentum tensor
$T^{{\overline {00}}}$ has the inertial mass tensor
$m^{ij}=T^{{\overline {00}}}\delta^{ij}
+T^{{\overline {ij}}}$ [12].

Unfortunately, at present, there is not yet a direct calculation
of inertial mass of a Casimir cavity in 4 dimensions; nevertheless
direct calculations of inertial mass on a two-dimensional space-time
[13] are in agreement with our calculation if ``extended'' to
four dimensions. Furthermore, it is very important to note that,
when considering the total force acting on the real cavity, which
is an isolated system, the contribution to the force resulting
from the spatial part of the energy-momentum tensor is balanced by
the contribution of the mechanical energy-momentum tensor, and hence
should not be considered for experimental evaluation. The resulting
force is then the Newtonian force on the sum of the rest
Casimir energy and rest mechanical mass: the contribution of
vacuum fluctuations leads to a gravitational
push on the Casimir apparatus expressed by the formula
\begin{equation}
{\vec F} \cong {\pi^{2}L^{2}{\hbar}c \over 720 a^{3}}
{g\over c^{2}}{\vec e}_{r},
\end{equation}
which should be tested against observation. As far as we can see,
our calculation suggests that the electromagnetic vacuum
state in a weak gravitational field is red-shifted, possibly adding
evidence in favour of virtual quanta being able to gravitate,
a non-trivial property on which no universal agreement has
been reached in the literature [8,9], as we already noticed in
the introduction.
To further appreciate this point,
we now find it appropriate and helpful to show that
the same result can be derived from a mode-by-mode analysis.
Recall indeed that in Minkowski space-time
the zero-point energy of the system can be evaluated, in the case
of perfect conductors, as
\begin{equation}
U={\hbar c L^{2}\over 2}\sum_{n=-\infty}^{\infty}
\int {d^{2}k \over (2\pi)^{2}}\sqrt{k^{2}+
\left({n \pi \over a}\right)^{2}},
\end{equation}
where we have considered the normal modes labelled by the
integer $n$ and the transverse momentum $k$. On performing the
integral by dimensional regularization, the energy takes the
well known Casimir expression
\begin{equation}
U_{\rm reg}=-{\pi^{2}L^{2}{\hbar}c \over 720 a^{3}},
\end{equation}
where the final result is independent of the particular
regularization method.

Consider next the cavity at rest in a Schwarzschild geometry.
On assuming that virtual quanta do gravitate, and bearing in mind
that the modes remain unchanged, the
energy of each mode is red-shifted by the factor
$\sqrt{g_{00}}=\sqrt{1-{\alpha \over r}}$, with $\alpha \equiv
{2GM\over c^{2}}$. Hence the total energy can be written as
(the suffix $S$ referring to Schwarzschild)
\begin{equation}
U_{S}={\hbar c L^{2}\over 2}\sum_{n=-\infty}^{\infty}
\int {d^{2}k \over (2\pi)^{2}}\sqrt{\left(k^{2}+
\left({n \pi \over a}\right)^{2}\right)(\sqrt{g_{00}})^{2}}.
\end{equation}
On performing dimensional regularization one finds therefore
\begin{equation}
(U_{S})_{\rm reg}=(g_{00})^{1\over 2}U_{\rm reg}
=-\left(1-{\alpha \over r}\right)^{1\over 2}
{\pi^{2}L^{2}{\hbar}c \over 720 a^{3}}.
\end{equation}
Now we assume that minus the gradient of $(U_{S})_{\rm reg}$
with respect to $r$ yields the force exerted by the gravitational
field on the rigid Casimir cavity. If $r >> \alpha$, we find
the result in Eq. (9) by working at the same level of approximation.
Higher order contributions are of order $\alpha^{2}r^{-3}$.

\section*{Problems of experimental verification}

In considering the possibility of experimental verification of the
extremely small forces linked to this effect we point out that
such measurements cannot be performed statically; this would make
it necessary to compare the weight of the assembled cavity with
the sum of the weights of its individual parts, a measure
impossible to perform. On the contrary, the measurements we are
interested in should be performed dynamically, by modulating the
force in a known way; the effect will be detected if the
modulation signal will be higher than the sensitivity of the
detector. In this spirit we focus on the sensitivity reached by
the present technology in detection of very small forces on a
macroscopic body, on earth, paying particular attention to
detectors of the extremely small forces induced by a gravitational
wave. As an example, gravitational wave signals $h$ of order
$\approx 10^{-25}$, corresponding to forces of magnitude $ \approx
5 \cdot 10^{-17} N$ at frequency of few tens of Hz, are expected
to be detected with the Virgo gravitational wave detector
presently under construction [14,15] after a month of
integration time.

In the course of studying experimental possibilities of
verification of the force on a rigid Casimir cavity, we  evaluate
this force on a macroscopic body, having essentially the  same
dimensions of mirrors for gravitational wave detection and
obtained through a multi-layer sedimentation by a series of rigid
cavities. Each rigid cavity consists of two thin metallic disks,
of thickness of order 100 nm [16] separated by a dielectric
material, inserted to maintain the cavity sufficiently rigid:
introduction of a dielectric is equivalent to enlarging the
optical path length by the refractive index $n$ $(a \mapsto na)$.
By virtue of presently low costs and facility of sedimentation,
and low absorption in a wide range of frequencies [17],
$SiO_{2}$ can be an efficient dielectric material.

{}From an experimental point of view we point out that the Casimir
force has so far been tested  down to a distance $a$ of about $60$
nm, corresponding to a frequency $\nu_{\rm min}$ of the
fundamental mode equal to $2.5 \cdot 10^{15} Hz$. This limit
results from the difficulty to control the distance between two
separate bodies, as in the case of measurements of the Casimir
pressure. As stated before, in our rigid case, present
technologies allow for cavities with much thinner separations
between the metallic plates, of the order of few nanometers. At
distances of order 10 nm, finite conductivity  and dielectric
absorption are expected to play an important role in decreasing
the effective Casimir pressure, with respect to the case of
perfect mirrors [18,19]. In this paper we discuss experimental
problems by relying  on present  technological resources,
considering cavities with plates' separation of 5 nm and
estimating the effect of finite conductivity by considering the
numerical results of Ref. [18]; for a separation of 6.5 nm this
corresponds to a decreasing factor $ \eta $  of about $7 \cdot
10^{-2}$ for Al. Moreover, to increase the total force and obtain
macroscopic dimensions, $N_{l} = 10^{6}$ layers can be used, each
having a diameter of 35 cm, and thickness of 100 nm, for a total
thickness of about 10 cm. Last, one should also consider
corrections resulting from finite temperature and roughness of the
surfaces, although one might hope to minimize at least the former
by working at low temperatures.

With these figures, the total force ${\vec F}^{T}$ acting on the
body can be calculated with the help of Eq. (9), modified to take
into account the refractive index $n$ for $SiO_{2}$, the
decreasing factor $\eta$, the area A of disk-shape plates, and the
$N_{l}$ layers:
\begin{equation}
{\vec F}^{T} \approx \eta N_{l} \frac{A \pi^{2} \hbar c}
{720(na)^{3}} \frac{ g}{c^{2}}{\vec e}_{r} .
\end{equation}
This formula describes a static effect, while the need for a
feasible experiment makes it mandatory to modulate the force, and
various experimental possibilities [20-22] are currently under
study. In particular, we are investigating the
possibility of modulating $\eta$, by varying the temperature, so
as to achieve a periodic transition from conductor to
superconductor regime. By doing so one can obtain $\eta_{\rm max}$
of order $5 \cdot 10^{-1}$ [23], and the magnitude of the force at the
modulation frequency can reach $10^{-14} \; N$. Even though such a
force is apparently more than two orders of magnitude larger than
the force which the Virgo gravitational antenna is expected to
detect, we should consider that the signal there is at
reasonable high frequency (some tens of Hz), while our calculated
signal remains at lower frequencies, i.e. tens of mHz. Moreover,
the technical problems resulting from stress induced in changing
temperature require careful consideration before saying that
modulation is feasible.

\section*{Conclusions}
The relation of the Casimir energy with the geometry of bounding
surfaces has been properly understood only very recently, thanks
to the outstanding work in Ref. [24], and another open problem
of modern physics, i.e. the cosmological constant problem,
results from calculations
which rely on the application of the equivalence principle to
vacuum energy [7], and this adds interest to our calculations,
that we have performed by focusing on a Casimir apparatus.

Our original contributions are given by the evaluation of the
force acting on a macroscopic body which mimics the rigid
Casimir cavity, and by a detailed estimate of the expected order
of magnitude of such a force.

As far as we can see, there is room left for an assessment of our
investigation, i.e.: (i) how to make sure that the cavity is
sufficiently rigid; (ii) the task of verifying that corrugations
or yet other defects do not affect substantially our estimates;
(iii) how to perform signal modulation, which is still beyond our
reach.

Our result expressed by Eq. (9), according to which the
gravitational field on a rigid  Casimir cavity gives rise to a net
force, whose direction is opposite to the one of gravitational
acceleration, is simple but non-trivial,
because it is a possible first
step towards testing the equivalence principle in configurations
involving the regularized zero-point energy of the electromagnetic
field. The effect can be understood in terms of the change in
rest mass because the system is more tightly bound as the cavity
walls are brought closer together. Interestingly, we therefore end up
by suggesting that a very tiny force produced by variations of rest
mass is detectable under suitable conditions, whereas standard
estimates of the change in rest mass for chemical reactions show
that this is unlikely to be the case [25].

The experimental verification of the calculated force,
which was our main concern,
might be performed if the problem of signal modulation could
be solved. On the other hand, in the authors' opinion,
the order of magnitude
of the calculated force can be already of interest
to demonstrate that experiments involving effects of gravitation
and vacuum fluctuations are not far from what can be obtained with
the help of present technological resources.
\section*{Acknowledgments}
The INFN financial support is gratefully acknowledged. The work of
G. Esposito has been partially supported by the Progetto di
Ricerca di Interesse Nazionale {\it SINTESI 2000}. We are indebted
to Bernard Kay and Serge Reynaud for correspondence,
and to Giuseppe Marmo for conversations. More
important, we acknowledge the lovely support of Cristina, Michela
and Patrizia.
\newpage
\section*{{\bfseries References}}

\noindent [1] B.S. DeWitt, Phys. Rep. 19 (1975) 295; M. Bordag, U.
Mohideen, V.M. Mostepanenko, Phys. Rep. 353 (2001) 1.

\noindent [2] M. Bordag, B. Geyer, G.L. Klimchitskaya, V.M.
Mostepanenko, Phys. Rev. D 60 (1999) 055004.

\noindent [3] K. Milton, Dimensional and Dynamical Aspects of the
Casimir Effect: Understanding the Reality and Significance of
Vacuum Energy (HEP-TH 0009173).

\noindent [4] S.A. Fulling, Aspects of Quantum Field Theory in
Curved Space-Time, Cambridge University Press, Cambridge, 1989.

\noindent [5] S.K. Lamoreaux, Phys. Rev. Lett. 83 (1999) 3340; U.
Mohideen, A. Roy, Phys. Rev. Lett. 81 (1998) 4549.

\noindent [6]
H.B. Chan, V.A.
Aksyuk, R.N. Kleiman, D.J. Bishop, F. Capasso, Science,
291 (2001) 1941.

\noindent [7] S. Weinberg, Rev. Mod. Phys. 61 (1989) 1.

\noindent [8] R.P. Feynman, A.R. Hibbs, Quantum Mechanics and
Path Integrals, McGraw Hill, New York, 1965, pp. 244--246;
for an up-to-date presentation, with references, see
P.S. Wesson, Astrophys. J. 378 (1991) 466.

\noindent [9] D.W. Sciama, in: The Philosophy of Vacuum, eds.
S. Saunders and H.R. Brown, Clarendon, Oxford, p. 137.

\noindent [10] C. Moller, The Theory of Relativity,
Oxford University Press, Oxford, 1972.

\noindent [11] L.S. Brown, G.J. Maclay, Phys. Rev. 184 (1969) 1272.

\noindent [12] C.W. Misner, K.S. Thorne, J.A. Wheeler,
Gravitation, Freeman, New York, 1973.

\noindent [13] M.T. Jaekel, S. Reynaud, Jour. Phys. I 3 (1993) 1093.

\noindent [14] D. Blair, The Detection of Gravitational Waves,
Cambridge University Press, Cambridge, 1991.

\noindent [15] VIRGO Collaboration, Nucl. Instr. Meth. A 360
(1995) 258; VIRGO Collaboration, in: Verbier 2000, Cosmology and
Particle Physics, p. 138.

\noindent [16] M. Bordag, B. Geyer, G.L. Klimchitskaya, V.M.
Mostepanenko, Phys. Rev. D 62 (2000) 011701.

\noindent [17] R.A. Street, Hydrogenated Amorphous Sylicon,
Cambridge University Press, Cambridge, 1991; B. Chapman, Glow
Discharge Processes, John Wiley, New York, 1980.

\noindent [18] A. Lambrecht, S. Reynaud, Eur. Phys. J. D 8 (2000)
309.

\noindent [19] G. Jordan Maclay, Phys. Rev. A 61 (2000) 052110.

\noindent [20] S. Hunklinger, H. Geisselmann, W. Arnold,
Rev. Sci. Instrum. 43 (1972) 584.

\noindent [21] P.R. Saulson, Interferometric Gravitational
Wave Detectors, World Scientific, Singapore, 1994.

\noindent [22] R. Esquivel--Sirvent, C. Villareal, G.H.
Cocoletzi, Phys. Rev. A 64 (2001) 052108.

\noindent [23] E. Calloni, L. Di Fiore, G. Esposito, L. Milano,
L. Rosa, Int. J. Mod. Phys. A 17 (2002) 804.

\noindent [24] G. Barton, J. Phys. A 34 (2001) 4083.

\noindent [25] E.F. Taylor and J.A. Wheeler, Spacetime Physics:
Introduction to Special Relativity, Second Edition,
Freeman, New York, 1992.

\end{document}